\documentclass[pra,10pt,twocolumn,reprint,superscriptaddress,floatfix,showpacs]{revtex4-2}

\usepackage[utf8]{inputenc}
\usepackage{amsthm}
\usepackage[T1]{fontenc}     
\usepackage[british]{babel}  
\usepackage[sc,osf]{mathpazo}
\usepackage[scaled=0.86]{berasans}  
\usepackage[colorlinks=true, citecolor=blue, urlcolor=blue]{hyperref}  

\usepackage{graphicx} 
\usepackage{subfig}
\usepackage{bitset}
\usepackage[babel]{microtype}  
\usepackage{amsmath,amssymb,amsthm,bm,amsfonts,mathrsfs,bbm} 
\usepackage{xspace}  
\usepackage{pgfplots}
\usepackage{xcolor,colortbl}
\usepackage{float}
\usepackage{comment}
\def\ba{\begin{equation}}
	\def\ea{\end{equation}}
\def\bea{\begin{eqnarray}}
	\def\eea{\end{eqnarray}}
\def\ben{\begin{equation*}}
	\def\een{\end{equation*}}
\def\bean{\begin{eqnarray*}}
	\def\eean{\end{eqnarray*}}
\def\bma{\begin{mathletters}}
	\def\ema{\end{mathletters}}
\def\bi{\begin{itemize}}
	\def\ei{\end{itemize}}

\newcommand{\be}{\begin{equation}}
	\newcommand{\ee}{\end{equation}}

\newcommand{\kommentar}[1]{}

\newcommand{\forget}[1]{}

\newtheorem{theorem}{Theorem}

\begin{document}
	
	\title{Revealing Hidden Non $n$-Locality In $n$-Local Star Network}
	\author{Kaushiki Mukherjee}
	\email{kaushiki.wbes@gmail.com}
	\affiliation{Department of Mathematics, Government Girls' General Degree College, Ekbalpore, Kolkata-700023, India.}
	\author{Biswajit Paul}
	\email{biswajitbbkm01082019@gmail.com}
	\affiliation{Department of Mathematics, Balagarh Bijoy Krishna Mahavidyalaya, Hooghly, West Bengal, India.}
	
	\begin{abstract}
Keeping pace with technological advancement, in the past decade, use of scalable networks have extended the study of quantum non-classicality beyond the regime of Bell-CHSH nonlocality. Present work provide characterization of non $n$-locality that can be exploited by incorporating filtering operations in star-shaped $n$-local networks. This in turn provide a framework of sequential $n$-local networks capable of generating non $n$-local correlations by involving some suitable form of stochastic local operations assisted with classical communications(SLOCC). It is observed that for effectiveness of such sequential networks, Bell-CHSH nonlocality(upto SLOCC operations) of every individual two-qubit state, distributed in the network, is not mandatory. However, there does not exist any separable local filter, which when applied in $n$-local network involving only Bell local states(upto SLOCC operations), can reveal non $n$-locality. 
Interestingly, instances revealing advantage of non-separable mutli-qubit local filters over separable mutli-qubit local filters(by central node) are obtained. Such an advantage is attributable to the specific topology of star-shaped $n$-local networks and thus can never be reflected in Bell scenario.
\end{abstract}
	
	\maketitle
	
	\section{Introduction}\label{intro}
The notion of Bell nonlocality\cite{sh1,sh2} constitutes one of
	the most profound non-classical feature of quantum correlations. Such correlations, more commonly known as \textit{Bell nonlocal correlations}, are inexplicable by any local hidden variable(LHV) theory\cite{sh3,sh4}. Apart from foundational importance, study of Bell nonlocality finds multifaceted applications in practical tasks\cite{prac1,prac2}. 
\par Violation of some appropriate Bell’s inequality ensures nonlocality of corresponding  correlations. Nonlocality of correlations in turn guarantees entanglement of particles constituting the physical system from which the correlations are generated after suitable spatially separated measurements in usual $(n,m,k)$ scenario. The converse is not always true as there exist several Bell-local mixed entangled states\cite{sh4,hir1}. However, some of those entangled states can violate Bell’s inequality when subjected to suitable sequential measurement scenario. In context of analyzing role of sequential measurements set-ups for exploiting Bell nonlocality, the application of local filtering operations followed by local measurements by individual parties,
	is noteworthy. Precisely, some entangled states can violate Bell inequality only if those are subjected to suitable local filtering operations prior to usual local measurements by spatially separated parties. Such form of Bell nonlocality is usually referred to as \textit{hidden Bell nonlocality}\cite{gis2,pop1}. Research over the years have demonstrated and analyzed the notion of hidden nonlocality from multiple perspectives\cite{gis2,pop1,hir2,bp1}. For example, \cite{gis2,pop1} provided instances of Bell-CHSH local entangled states which exhibit hidden nonlocality. In \cite{hir2}, the authors have pointed out the possibility of generating nonlocal correlations even from entangled states admitting a LHV model under suitable
	measurement context(involving filtering operations). In \cite{rajs}, the authors have established a necessary and sufficient criterion to detect hidden nonlocality of arbitrary two-qubit state. In \cite{locacces}, the authors pointed out existence of two-qubit states whose hidden quantum correlations can be revealed only when both parties cooperate in applying local filters whereas no such non-classicality can be exploited when any one of the parties does not perform filtering operations. From broader perspectives, present work will be contributory in the direction of adding insights to hidden nonlocality theory pertaining to quantum systems beyond usual Bell scenario($(n,m,k)$ scenario). 
	\par Speaking of exploring quantum systems beyond usual Bell measurement contexts, study of quantum networks deserve special mention\cite{birev}. In recent past nonlocality of correlations have been studied rigorously in networks involving multiple independent quantum sources\cite{frtz1,BRAN,BRA,km1,km2,gis1,Tava,bilo2,lee,bilo3,bilo4,km4,km5,bilo5,bilo6,ejm,bilo7,nr1,nr2,nr3,nr4,km7,kau6,km8}. In any such network scenario, more commonly known as \textit{$n$-local networks}, each of the independent sources distributes physical states\cite{BRAN,BRA} to only a subset of all the parties involved therein. Due to the source independence assumption($n$-local assumption), quantum correlations generated in $n$-local networks exhibit some novel non-classical features that have no analogue in usual Bell scenario. For instance, one can get non $n$-local correlations across the entire network even in absence of direct common past in between all the parties. Also, non $n$-locality can be observed even when some of the parties perform single measurement\cite{gis1,bilo2}. This is in striking contrast to standard Bell scenario where presence of randomness in measurements choice by each part is necessary to detect Bell nonlocality. Apart from the foundational significance, quantum network nonlocality is emerging as an important non-classical resource in different information processing
	tasks\cite{sh5,sh6,sh7,sh8}.	
	 	\par Owing to compatibility of $n$-local networks to several experimental setups\cite{birev}, study of this type of networks has been made in different directions. Present work will also probe $n$-local networks with the motivation to analyze detectable non $n$-local network correlations obtained in sequential measurement contexts. To be precise, we will study revelation of non $n$-locality from two-qubit state on being subjected to suitable local filtering operations. The notion of hidden non $n$-locality was first introduced in \cite{kau6}. There the authors analyzed hidden non $n$-locality in linear $n$-local networks. In context of existing findings related to hidden non $n$-locality, it becomes imperative to explore the topic in non linear network topology. 
	\par Present discussion will consider sequential non-linear $n$-local network endowed with star topology. In usual star shaped $n$-local networks, there exists a closed form of the upper bound of the $n$-local inequality\cite{Tava}. A closed form of the existing $n$-local inequality will be formalized here in terms of the parameters characterizing the independent sources and the local filtering operations. Such a form of the upper bound will aid in exploiting instances of hidden non trilocality. In course of analyzing hidden non $n$-locality it becomes pertinent to characterize the two-qubit states distributed by the independent sources across the network. From viewpoint of characterization, some queries need to be addressed:
\begin{itemize}
		\item \textit{can hidden non $n$-local correlations be detected when all the two-qubit states, distributed by the sources, remain Bell-CHSH local even after application of local filtering operations?}
		\item \textit{contrary to above, for generating non $n$-locality, is it necessary for every source to distribute a hidden Bell-nonlocal state?}
\end{itemize}
Negative responses of both these queries will be provided here. 
In cases when only some of the parties are allowed to perform filtering operations, it becomes important to analyze whether non $n$-local correlations can still be observed or not. In this context, notion of \textit{locally accessible non $n$-locality} is introduced here. Interestingly, non trilocality is witnessed when either the extreme parties or the central party(not both simultaneously) perform suitable local filtering operations. From practical viewpoint, instances of hidden nontrilocality provided herein points out enhancement in robustness to noise of existing $n$-local inequality when used for detecting non $n$-locality in sequential $n$-local star network compared to non-sequential(usual) $n$-local star network.   
\par In  \cite{kau6} it was considered that each of the central parties can perform only separable form of local filters. However, as per the topology, the central node gets access to multiple qubits(getting a qubit from each source).
To this end an obvious query arises: \textit{can non-separable local filtering operations at the central node of the network provide advantage for exploiting hidden non $n$-locality?} Findings ensuring advantage of multi-qubit non-separable local filters over multi-qubit separable local filters, will be provided here.
\par Rest of the manuscript is organized as follows: in sec.\ref{pre1} the mathematical pre-requisites are discussed. Sequential network based protocol will be set up in sec.\ref{ress1} followed by characterization of hidden non $n$-locality in sec.\ref{ress2}. Local accessibility of hidden non $n$-locality is next discussed in sec.\ref{ress3}. Practical implementation of present work is discussed in \ref{ress4}. Discussion of non-separable filters is given in \ref{ress6}. Some concluding remarks are provided in sec.\ref{ress7}.

	\section{Preliminaries}\label{pre1}
	\subsection{Bloch Matrix Representation}\label{bloch}
	Let $\rho$ denote any two qubit state shared between two parties Alice and Bob. In terms of Bloch parameters $\rho $ can be written as:
	\begin{equation}\label{st4}
		\small{\rho}=\small{\frac{1}{4}(\mathbb{I}_{2\times2}+\vec{x}.\vec{\sigma}\otimes \mathbb{I}_2+\mathbb{I}_2\otimes \vec{y}.\vec{\sigma}+\sum_{k_1,k_2=1}^{3}r_{k_1k_2}\sigma_{k_1}\otimes\sigma_{k_2})},
	\end{equation}
	with $\vec{\sigma}$$=$$(\sigma_1,\sigma_2,\sigma_3), $ $\sigma_{i}$ denoting Pauli operators along $3$ mutually perpendicular directions($i$$=$$1,2,3$). $\vec{x}$$=$$(x_1,x_2,x_3)$ and $\vec{v}$$=$$(y_1,y_2,y_3)$ represent the local bloch vectors($\vec{x},\vec{y}$$\in$$\mathbb{R}^3$) of  Alice and Bob respectively with $|\vec{x}|,|\vec{y}|$$\leq$$1$ and $(r_{i,j})_{3\times3}$ denoting the correlation tensor $\mathcal{R}$(real matrix).
	$\mathcal{R}'$s components are given by $r_{ij}$$=$$\textmd{Tr}[\rho\,\sigma_{i}\otimes\sigma_{j}].$ \\
	$\mathcal{R}$ can be diagonalized by subjecting to suitable local unitary operations\cite{gam}:
	\begin{equation}\label{st41}
		\small{\rho}^{'}=\small{\frac{1}{4}(\mathbb{I}_{2\times2}+\vec{\mathfrak{a}}.\vec{\sigma}\otimes \mathbb{I}_2+\mathbb{I}_2\otimes \vec{\mathfrak{b}}.\vec{\sigma}+\sum_{j=1}^{3}W_{j}\sigma_{j}\otimes\sigma_{j})},
	\end{equation}
	Here the correlation tensor is given by $\textbf{W}$$=$$\textmd{diag}(W_{1},W_{2},W_{3}).$ $W_{1},W_{2},W_{3}$ are the eigenvalues of $\sqrt{\mathcal{R}^T\mathcal{R}},$ i.e., singular values of $\mathcal{R}$ arranged in descending order of magnitude, i.e., $W_{1}$$\geq$$W_{2}$$\geq$$W_{3}.$
	\subsection{Bell-Diagonal States}
	For any integer $i$$\geq$$ 1,$ let $\rho_i$ denote a  two-qubit Bell-diagonal state\cite{nie}:
	\begin{eqnarray}\label{belld1}
		\rho_i^{(\small{BD})}&=&w_{i1}|\psi^{+}\rangle\langle \psi^{+}|+w_{i2}|\psi^{-}\rangle\langle \psi^{-}|+w_{i3}|\phi^{+}\rangle\langle \phi^{+}\nonumber\\
		&+&w_{i4}|\phi^{-}\rangle\langle \phi^{-}|,\,\textmd{with}\nonumber\\
		&&0\leq w_{i1},w_{i2},w_{i3},w_{i4}\leq 1,\,\sum_{j=1}^{4}w_{ij}=1\nonumber\\
		&&|\psi^{\pm}\rangle\langle \psi^{\pm}|,|\phi^{\pm}\rangle\langle \phi^{\pm}|\textmd{ \small{denoting Bell states.}}
	\end{eqnarray}
	Bell-diagonal states(Eq.(\ref{belld1})) are Bell-CHSH local\cite{horo} if:
	\begin{eqnarray}\label{belld2}
		\textmd{Max}[\sqrt{E_{i1}},\sqrt{E_{i2}},\sqrt{E_{i3}}]&\leq&1,\,\textmd{with}\nonumber
		\\
		W_{i1}=(w_{i1} - w_{i2} + w_{i3} - w_{i4})^2 &+& (w_{i1} - w_{i2} - w_{i3} + w_{i4})^2,\nonumber\\
		W_{i2} =(w_{i1} - 
		w_{i2} + w_{i3} - w_{i4})^2 &+& (-w_{i1} - w_{i2} + w_{i3} + w_{i4})^2 \nonumber\\
		W_{i3}=	(w_{i1} - w_{i2} - w_{i3} + 
		w_{i4})^2 &+& (-w_{i1} - w_{i2} + w_{i3} + w_{i4})^2\nonumber\\
		&&
	\end{eqnarray}
	Any Bell-CHSH local Bell-diagonal state remains local even after application of local filtering operations\cite{verst}. So if Eq.(\ref{belld2}) holds, then corresponding Bell-diagonal state is Bell-CHSH local upto SLOCC. \\
	\subsection{Necessary $\&$ Sufficient Criterion of Hidden Bell-CHSH Nonlocality}
	Let $\rho$ denote any two qubits state shared between $A$ and $B.$ Let $F_{A}$ and $F_B$ denote any qubit filtering operation applied by $A$ and $B$ over their respective qubits. Filtered state $\rho^{'}$(say) takes the form:
	\begin{equation}\label{pref1}
		\rho^{'}=\frac{((F_A\otimes F_B).\rho.(F_A\otimes F_B)^{\dagger})}{\textmd{Tr}[(F_A\otimes F_B).\rho.(F_A\otimes F_B)^{\dagger}]}
	\end{equation}
	Term $\textmd{Tr}[(F_A\otimes F_B).\rho.(F_A\otimes F_B)^{\dagger}]$ is the probability of obtaining the filtered state $\rho^{'}$ from $\rho.$ This term is also  know as success probability of the filtering operation.\\
	$\rho$ is said to display hidden Bell-CHSH nonlocality if $\rho^{'}$ violates Bell-CHSH inequality for at least some filtering operations $F_A,F_B.$\\
	Let $Q$ denote the following matrix corresponding to an arbitrary two-qubit state $\rho$(say):
	\begin{equation}
		Q_{\rho}=[Q_{ij}]_{i,j=0,1,2,3}=\textmd{Tr}[\rho.(\sigma_i\otimes \sigma_j)]
	\end{equation}
	with $\sigma_0$ representing $2\times 2$ identity matrix.\\
	Let $M_{\rho}$ denote the matrix:
	\begin{eqnarray}
		M&=&U.Q.U.Q^{T}\,\,\textmd{where}\nonumber\\
		U&=&\textmd{diag}[1,-1,-1,-1]
	\end{eqnarray}
	$M$ is a $4\times 4$ matrix. Let $\mu_1$$\geq$$\mu_2$$\geq$$\mu_3$$\geq$$\mu_4$ denote ordered eigen values of $M.$\\
	$\rho$ shows hidden Bell-CHSH nonlocality if and only if the following criterion holds\cite{rajs}:
	\begin{equation}\label{rajc}
		\mu_2+\mu_3>\mu_1.	
	\end{equation}
	\subsection{Star Network\cite{Tava}}\label{star1}
	\begin{center}
		\begin{figure}
			\includegraphics[width=3.5in]{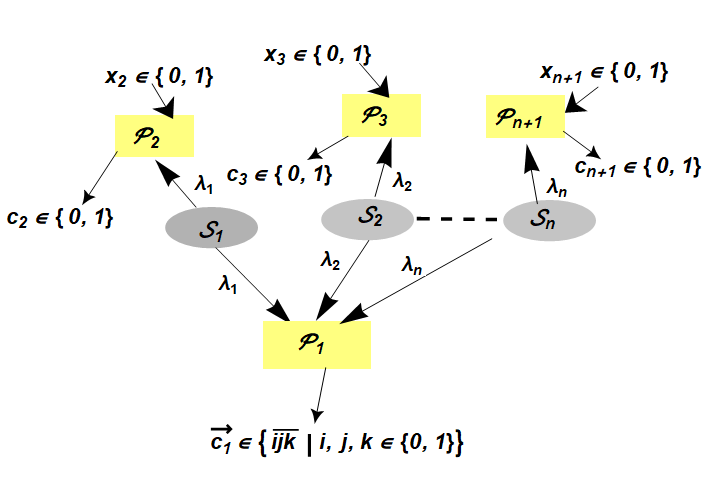} \\
			\caption{\emph{Schematic representation of $n$-local star network. }}
			\label{star}
		\end{figure}
	\end{center}
	Star network is a non-linear $n$-local network consisting of a single central party $\mathcal{P}_1$ and $n$ edge(extreme) parties $\mathcal{P}_2,...,\mathcal{P}_{n+1}$ (see Fig.\ref{star}). Let the network be denoted by $\mathcal{N}_{n-star}.$\\
	Each of the edge parties receives one particle from a source. Each of the sources $\mathcal{S}_i$ is characterized by variable $\lambda_i(i$$=$$1,2,...,n)$ The sources being independent, joint distribution of the variables $\lambda_1,...,\lambda_n$ is factorizable:
	\begin{equation}\label{tr1}
		q(\lambda_1,\lambda_2...\lambda_n)=\Pi_{i=1}^n q_i(\lambda_i),
	\end{equation}
	with $q_i$ denotes the normalized distribution of $\lambda_i,\forall i.$ Source independence condition(Eq.(\ref{tr1})) represents the $n$-local constraint\cite{BRAN}. \\
	Each of the edge parties $\textbf{P}_i$ chooses to perform from any one of two possible measurements($x_i$$\in$$\{0,1\}$) whereas the central party performs a fixed measurement. $n+1$ partite measurement correlations are local if:
	\begin{eqnarray}\label{tr2}
		&&\small{p(\vec{\mathfrak{c}}_1,c_2,c_3...,c_{n+1}|x_2,x_3,...,,x_{n+1})}=\nonumber\\
		&&\int_{\Lambda_1}\int_{\Lambda_2}...\int_{\Lambda_n}
		d\lambda_1d\lambda_2...d\lambda_n\,q(\lambda_1,\lambda_2,...\lambda_n) \mathcal{P} ,\,\textmd{with}\nonumber\\
		&&\mathcal{P}=p(\vec{\mathfrak{c}}_1|\lambda_1,...,\lambda_n)\Pi_{i=1}^{n} p(c_{i+1}|x_{i+1},\lambda_i)\nonumber\\
		&&
	\end{eqnarray}
	$n+1$-partite correlations satisfying both Eqs.(\ref{tr1},\ref{tr2}) are termed as $n$-local correlations. Hence, any set of correlations that do not satisfy Eqs.(\ref{tr1},\ref{tr2}) simultaneously, are termed as non $n$-local\cite{BRAN}.\\ 
	
	The existing $n$-local inequality for $\mathcal{N}_{n-star}$ is given by\cite{bilo2}:
	\begin{eqnarray}\label{ineqs}
		&& \sum_{i=1}^{2^{n-1}}|J_{i}|^{\frac{1}{n}}\leq  2^{n-2},\,  \textmd{where}\\
		&& J_i=\frac{1}{2^n}\sum_{x_2,...,x_{n+1}}(-1)^{g_i(x_2,...,x_{n+1})}\langle A_{(1)}^{(i)}A_{x_2}^{(2)}...A_{x_{n+1}}^{(n+1)}\rangle\nonumber\\
		&&\langle A_{(1)}^{(i)}A_{x_2}^{(2)}...A_{x_{n+1}}^{(n+1)}\rangle=\sum_{\mathcal{D}_1}(-1)^{\tilde{\mathfrak{c}}_1^{(i)}+\textbf{c}_2+...
			+\textbf{c}_{n+1}}N_1,\nonumber\\
		&& \textmd{\small{where}}\,N_1=\small{p(\overline{\textbf{c}}_1,
			\textbf{c}_2,...,\textbf{c}_{n+1}
			|x_2,...,x_{n+1})}\,\textmd{\small{and}}\nonumber\\
		&&\mathcal{D}_1=\{\textbf{c}_{11},....,\textbf{c}_{12^n},\textbf{c}_2,....
		,\textbf{c}_{n+1}\}\nonumber\\
	\end{eqnarray}
	In Eq.(\ref{ineqs}), $\forall i$$=1,2,...,2^{n-1},$ $\tilde{\mathfrak{c}}_1^{(i)}$ represents an output bit generated by classical post-processing of the raw output string $\overline{\textbf{c}}_1$$=$$(\textbf{c}_{11},....,\textbf{c}_{12^n})$ of $\mathcal{P}_1.$ In Eq.(\ref{ineqs}), $\forall i,\, g_i$ are functions of the input variables $x_2,...,x_{n+1}$ of the extreme parties\cite{Tava}. Each $g_i$ contains an even number of $x_2,...,x_{n+1}.$ For $n$$=$$3,4$, classical post-processed bits $\tilde{\mathfrak{c}_1^{(i)}}$ from the output string $\overline{\textbf{c}_1}$ and corresponding functions $g_i(x_2,x_3,...,x_{n+1})$ are specified in Table.\ref{table:ta6}.
	Let each of the sources distribute an arbitrary two qubit state (Eq.\ref{st41}). Under above measurement settings, the upper bound of Eq.(\ref{ineqs}) is given by $	B_{n-star}$\cite{bilo5} where: 
	\begin{equation}\label{boundstar}
		B_{n-star}=	2\sqrt{(\Pi_{i=1}^n  W_{i1})^{\frac{2}{n}}+(\Pi_{i=1}^nW_{i2})^{\frac{2}{n}}}.
	\end{equation}
	$n$-local inequality (Eq.(\ref{ineqs})) is thus violated if:
	\begin{equation}\label{up211}
		B_{n-star}>2.
	\end{equation}
	Hence, in case Eq.(\ref{up211}) is violated, the corresponding network correlations turn out to be non $n$-local as per the $n$-local inequality (Eq.(\ref{ineqs})). 
	\begin{table}
		\caption{Table displays the details of the classically post-processed bits $\tilde{\mathfrak{c}}_1^{(i)}$ and also the functions $g_i(x_2,x_3,...,x_{n+1})$ for $n$$=$$3,4$ in the $n$-local inequality(Eq.(\ref{ineqs})).}
		\begin{center}
			\begin{tabular}{|c|c|c|}
				\hline
				$n$&$\tilde{\mathfrak{c}}_1^{(i)}$ &$g_i(x_2,x_3,...,x_{n+1})$  \\
				\hline
				$3$&$\tilde{\mathfrak{c}}_1^{(1)}$$=$$\textbf{c}_{11}$,
				$\tilde{\mathfrak{c}}_1^{(2)}$$=$$\textbf{c}_{11}$$\oplus$$\textbf{c}_{12}$$\oplus$$1$&$g_1$$=$$0,$
				$g_2$$=$$x_2$$+$$x_3$\\
				
				&$\tilde{\mathfrak{c}}_1^{(3)}$$=$$\textbf{c}_{11}$$\oplus$$\textbf{c}_{13}$$\oplus$$1$
				&$g_3$$=$$x_2$$+$$x_4$\\
				&$\tilde{\mathfrak{c}}_1^{(4)}$$=$$\textbf{c}_{11}$$\oplus$$\textbf{c}_{12}$$\oplus$$\textbf{c}_{13}
				$$\oplus$$1$&$g_4$$=$$x_3$$+$$x_4$\\
				&\,&\,\\
				\hline
				&$\tilde{\mathfrak{c}}_1^{(1)}$$=$$\textbf{c}_{11}$,
				$\tilde{\mathfrak{c}}_1^{(2)}$$=$$\textbf{c}_{11}$$\oplus$$\textbf{c}_{12}$$\oplus$$1$&
				$g_1$$=$$0,$
				$g_2$$=$$x_2$$+$$x_3$\\
				
				$4$&$\tilde{\mathfrak{c}}_1^{(3)}$$=$$\textbf{c}_{11}$$\oplus$$\textbf{c}_{13}$$\oplus$$1$
				&$g_3$$=$$x_2$$+$$x_4$\\
				
				&$\tilde{\mathfrak{c}}_1^{(4)}$$=$$\textbf{c}_{11}$$\oplus$$\textbf{c}_{14}$$\oplus$$1$
				&$g_4$$=$$x_2$$+$$x_5$\\
				
				&$\tilde{\mathfrak{c}}_1^{(5)}$$=$$\textbf{c}_{11}$$\oplus$$\textbf{c}_{12}$$
				\oplus$$\textbf{c}_{13}$$\oplus$$1$
				&$g_5$$=$$x_3$$+$$x_4$\\
				
				&$\tilde{\mathfrak{c}}_1^{(6)}$$=$$\textbf{c}_{11}$$\oplus$$\textbf{c}_{12}$$
				\oplus$$\textbf{c}_{14}$$\oplus$$1$
				&$g_6$$=$$x_3$$+$$x_5$\\
				
				&$\tilde{\mathfrak{c}}_1^{(7)}$$=$$\textbf{c}_{11}$$\oplus$$\textbf{c}_{14}$$
				\oplus$$\textbf{c}_{13}$$\oplus$$1$
				&$g_7$$=$$x_4$$+$$x_5$\\
				
				&$\tilde{\mathfrak{c}}_1^{(8)}$$=$$\textbf{c}_{11}$$\oplus$$\textbf{c}_{12}$$
				\oplus$$\textbf{c}_{13}$$\oplus$$\textbf{c}_{14}$
				&$g_8$$=$$x_2$$+$$x_3$$+$$x_4$$+$$x_5$\\
				\hline
			\end{tabular}
		\end{center}
		\label{table:ta6}
	\end{table}
	
	\section{Sequential Star Shaped $n$-local Network}\label{ress1}
	\begin{center}
		\begin{figure}
			\includegraphics[width=3.5in]{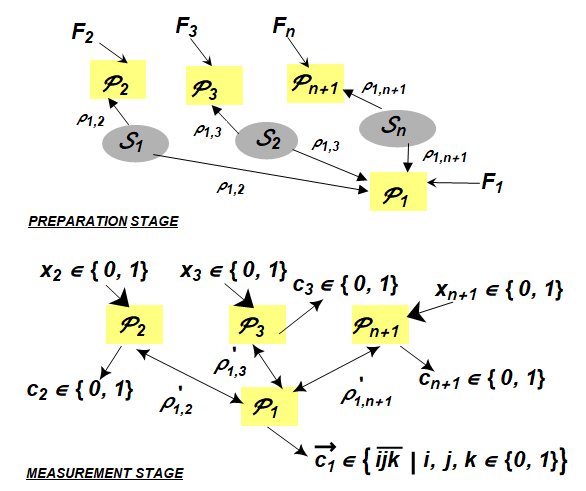} \\
			\caption{\emph{Schematic representation of the sequential star shaped $n$-local network protocol. }}
			\label{fig2}
		\end{figure}
	\end{center}
	The protocol is based on a $n$-local network in star configuration. Let the parties be allowed to perform local filtering operations before performing local measurements. Precisely, the entire protocol (see Fig.\ref{fig2})  comprises of two phases:
	\begin{enumerate}
		\item  \textit{Preparation Phase}
		\item \textit{Measurement Phase}
	\end{enumerate}
	\textit{Preparation Phase:} As in any $n$-local star network (sec.\ref{pre1}), let each of $n$ sources $\textbf{S}_i$ distribute a two-qubit state $\rho_{1,i}$ between $\textbf{P}_1$ and $\textbf{P}_{i}(i$$=$$2,...,n+1).$ Overall state of the physical systems shared by all the parties across the entire network thus takes the form:
	\begin{equation}\label{fil4}
		\rho_{\small{initial}}=\otimes_{i=2}^{n+1}\rho_{1,i}
	\end{equation}
	On receiving the particles, let each of the edge parties $\textbf{P}_{i}(i$$=$$2,...,n+1)$ perform $2$ dimensional local filtering operation $\textbf{F}_i$(say) on its respective subsystem. The central party $\textbf{P}_1$ performs a two dimensional local filter on each of the $n$ qubits received from the sources. $\forall i$$=$$2,3,...,n+1,$ let $\textbf{F}_1^{(i)}$ denote filtering operation that $\textbf{P}_1$ applies on qubit received from  $i^{th}$ source $\mathcal{S}_i.$ So, here, the overall filtering $\mathfrak{F}$(say) that the central party applies on the joint state of all its $n$ qubits(received from $n$ independent sources), is in separable form:
	\begin{equation}\label{cenfil}
		\mathfrak{F}=\otimes_{i=1}^{n}\textbf{F}_1^{(i)}
	\end{equation} 
	After all the parties perform filtering, the filtered state shared across network takes the form:
	\begin{eqnarray}\label{fil7}
		\rho^{(f)}&=&\textbf{N}(\mathfrak{F}\otimes_{i=2}^{n+1}  \textbf{F}_{i}).\rho_{\small{initial}}.(\mathfrak{F}\otimes_{i=2}^{n+1}  \textbf{F}_{i})^{\dagger}\nonumber\\
		\textmd{where}\,\textbf{N}&=&\frac{1}{\textmd{Tr}((\mathfrak{F}\otimes_{i=2}^{n+1}  \textbf{F}_{i}).\rho_{\small{initial}}.(\mathfrak{F}\otimes_{i=2}^{n+1}  \textbf{F}_{i})^{\dagger})}
	\end{eqnarray}
	$\frac{1}{\textbf{N}}$ denotes the probability of obtaining $\rho^{(f)}$ and is referred to as \textit{success probability} of the protocol. 
	\textit{Measurement Phase:} In this step each of the parties now performs local measurements on their respective share of particles forming the state $\rho^{(f)}.$ Measurement context is same as in the usual $n$-local star network scenario (sec.\ref{pre1}). To be precise, each of the edge parties $\textbf{P}_i$ performs  projective measurements $(\textbf{M}_{i0},\textbf{M}_{i1})$ along any one of two arbitrary directions: $\{\vec{m}_{i0}.\vec{\sigma},\vec{m}_{i1}.\vec{\sigma}\}$ for $\textbf{P}_i$($\forall i$$=$$2,3,...,n+1$) with $\vec{m}_{i0},\vec{m}_{i1}$$\in$$ \mathbb{R}^3.$ \\
	Central party $\textbf{P}_1$ performs projective measurement in GHZ basis\cite{Tava}.
	Local measurement statistics are then used to test a violation of the $n$-local inequality (Eq.(\ref{ineqs})).\\
	\par The measurement settings considered in this step of the protocol are same as that considered for usual star  $n$-local network\cite{bilo5}. It may be noted that it is the preparation phase for which the current scenario differs from the usual $n$-local star network scenario. In the usual scenario, there is no preparation phase. $\rho_{\small{initial}}$ is the overall state used in the measurement step of the usual scenario in contrast to the post-selected state $\rho^{(f)}$ in the sequential scenario.\\
	\par Let $\mathcal{N}_{n-star}^{(seq)}$ denote the sequential $n$-local star network scenario involved in our protocol. We next proceed to characterize the non $n$-locality of the $n+1$-partite correlations generated in $\mathcal{N}_{n-star}^{(seq)}$.
	\section{Characterizing Hidden Non $n$-local Correlations in $\mathcal{N}_{n-star}^{(seq)}$}\label{ress2}
	In this section, we will explore the effectiveness(if any) of applying local filters in context of detecting non $n$-local correlations. For that we first formally define hidden non $n$-local correlations in sequential $n$-local star network(with each party applying local filters on its respective physical systems) as follows:\\
	\textbf{Definition.1:}\textit{With each of the edge parties performing
		projective measurements in anyone of two possible directions
		and the central party performing single projective
		measurement in $n$-partite GHZ basis, under the n-local constraint(Eq.(\ref{tr1})), if $n+1$-partite measurement correlations obtained in the sequential
		star n-local network cannot be decomposed in the form given by
		Eq.(\ref{tr2}), then such correlations are said to be hidden non n-local
		correlations and the corresponding notion of nonlocality is referred to as hidden non n-locality.}\\
	In $\mathcal{N}_{n-star}^{(seq)}$, the upper bound of the $n$-local inequality(Eq.(\ref{ineqs})) can be expressed in terms of the Bloch parameters of the states distributed by the sources along with the parameters characterizing the single qubit filters applied by the parties on their respective subsystems. Precisely the following result holds:
	\begin{theorem}\label{theo1}
		\textit{In the sequential $n$-local star network, upper bound of $n$-local inequality(Eq.(\ref{ineqs})) is given by:}
		\begin{eqnarray}\label{fil7ii}
			B_{n-star}^{(seq)}&=&2\sqrt{\Pi_{i=1}^n  (W_{i1}^{(f)})^{\frac{2}{n}}+\Pi_{i=1}^n(W_{i2}^{(f)})^{\frac{2}{n}}},\\
			\forall i,\, W_{i1}^{(f)}&\geq& W_{i2}^{(f)}\geq W_{i3}^{(f)} \textmd{denote the ordered singular  }\nonumber	\\
			&& \textmd{values of the correlation tensor of the state}\nonumber\\
			&& \frac{\textbf{F}_{i+1}\otimes \textbf{F}_1^{(i)}.\rho_{1,i}.(\textbf{F}_{i+1}\otimes \textbf{F}_1^{(i)})^{\dagger}}{\textmd{Tr}[\textbf{F}_{i+1}\otimes \textbf{F}_1^{(i)}.\rho_{1,i}.(\textbf{F}_{i+1}\otimes \textbf{F}_1^{(i)})^{\dagger}]}
		\end{eqnarray}
	\end{theorem}
	\textit{Proof:} See Appendix.A.\\
	Non $n$-locality is detected in the sequential network if:
	\begin{equation}\label{up211f}
		B_{n-star}^{(seq)}>2
	\end{equation}
	Above theorem indicates that violation of Eq.(\ref{ineqs}) depends on the filter parameters and also both local Bloch vector parameters and correlation tensors of the states used in the network. Hidden non $n$-locality is revealed in the network if Eq.(\ref{up211}) is violated but Eq.(\ref{up211f}) is satisfied. 
	\par We next state a no-go result in the context of hidden non $n$-locality. Analyzing relaxation of one or more constraints of the result will thus provide directions for illustrating possibilities of detecting non $n$-locality. \\
	\begin{theorem}\label{theo2}
		\textit{In the sequential $n$-local star network, violation of $n$-local inequality(Eq.(\ref{ineqs})) is impossible if each of the independent sources distributes a two-qubit Bell-CHSH local state that remains Bell-CHSH local even after application of any local qubit filtering operations}.
	\end{theorem}
	\textit{Proof:} See Appendix.B.\\
	Theorem.\ref{theo2} serves as a sufficient criterion for ensuring impossibility of detecting any non $n$-local phenomenon in a $n$-local star network. This is because, distribution of two-qubit Bell-CHSH local(upto local filtering) state by each of $n$ sources suffices to guarantee no generation of detectable non $n$-local network correlations. \\
	Let us now relax the requirements for above no-go result in order to illustrate hidden non $n$-locality. For illustration purpose we consider trilocal sequential network($n$$=$$3$). Also let the forms of the local filters of the edge parties $\textbf{P}_i$ be:
	\begin{eqnarray}\label{filters1}
		\textbf{F}_i&=&\epsilon_i|0\rangle\langle 0|+|1\rangle\langle 1|,\,i=2,3,4\nonumber\\
	\end{eqnarray}
	and that applied by the central party $\textbf{P}_1$ be:
	\begin{eqnarray}\label{filters2}
		\textbf{F}_1^{(j)}&=&\epsilon_1^{(j)}|0\rangle\langle 0|+|1\rangle\langle 1|\nonumber\\
		\textmd{So, }\mathfrak{F}&=&\otimes_{j=1}^3(\epsilon_1^{(j)}|0\rangle\langle 0|+|1\rangle\langle 1|)
	\end{eqnarray}
	
	\subsection{Hidden Non Trilocality}
	First we consider the case where any one of the sources in the trilocal network distribute two qubit state that are Bell-CHSH local upto SLOCC or after application of any local qubit filtering operations. Remaining source will distribute two-qubit states that can display hidden Bell nonlocality.
	\subsubsection{$\mathcal{N}_{n-star}^{(seq)}$ Involving Two Hidden Nonlocal States}\label{hidex1}
\begin{center}
		\begin{figure}
			\includegraphics[width=3.4in]{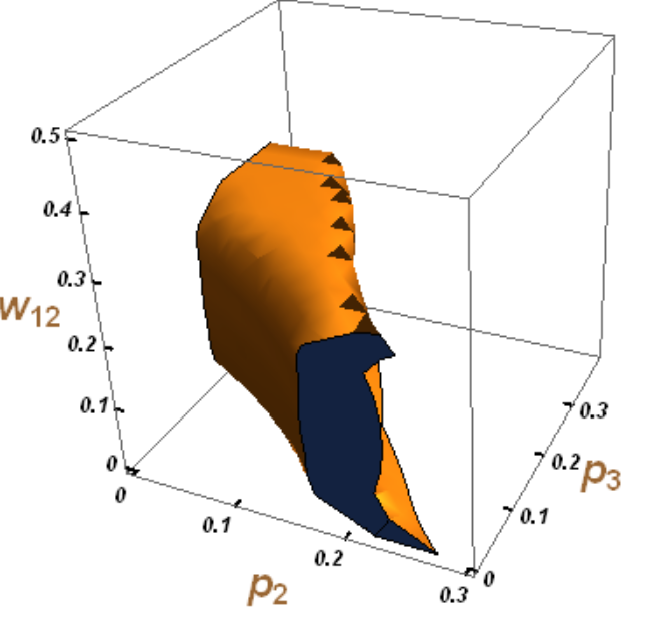} 
			\caption{\emph{Shaded region indicates parameters $w_{12},p_2,p_3$ of $\rho_1^{(\small{BD})},\rho_2^{(\small{H})}$ and $\rho_3^{\small{H}}$ respectively for which hidden non-trilocality is detected via our sequential protocol with approximately $32\%$ probability of success. The specifications of other state parameters and filtering parameters(Eqs.(\ref{filters1},\ref{filters2})) are as follows:
					$w_{11}$$=$$ 0.05,$	$w_{13}$$=$$ 0.639$,	$\theta_{2}$$=$$\theta_3$$=$$0.458,$ $\epsilon_2$$=$$0.99,$ $\epsilon_3$$=$$0.902,$ $\epsilon_4$$=$$0.998,$ $\epsilon_1^{(1)}$$=$$0.98,$ $\epsilon_1^{(2)}$$=$$0.5569,$  $\epsilon_1^{(3)}$$=$$0.98.$ Out of $3$ states involved, only $\rho_2^{(H)},\rho_3^{(H)}$ display  hidden Bell nonlocality. }}
			\label{2g1bs}
		\end{figure}
	\end{center}

	Let one source($\mathcal{S}_1$,say) distribute Bell-CHSH local Bell-diagonal state, i.e., $\rho_{1,1}$$=$$\rho_1^{(\small{BD})}$ satisfies Eq.(\ref{belld2}). So, $\mathcal{S}_1$ distribute state that do not display any hidden nonlocality(violates Eq.(\ref{rajc})).    \\
	Let each of $\mathcal{S}_2,\mathcal{S}_3$ distribute a state from the following two-qubit entangled family of states\cite{gr1}:
	\begin{eqnarray}\label{horodeck1}
		\rho_{1,i}=\rho_i^{\small{H}}&=&(1-p_i) (\sin\theta_i|01\rangle+ \cos\theta_i|10\rangle)\cdot(\sin\theta_i\langle 01|+\nonumber\\
		\cos\theta_i\langle 10|)&+&p_i|00\rangle\langle00|,\small{\textmd{with}}\,\,	p_i\in [0,1]\\
		&&\theta_i\in[0,\frac{\pi}{4}]\,\,i=2,3.
	\end{eqnarray}
	For any value of $p_i,\theta_i$ $\rho_i^{\small{H}}$ display hidden nonlocality, i.e., satisfy Eq.(\ref{rajc}). For some members of local $\rho_1^{\small{BD}}$ and hidden nonlocal $\rho_2^{\small{H}}$, $\rho_3^{\small{H}},$ hidden non-trilocal correlations are generated in the network under application of suitable local filters by the parties(see Fig.\ref{2g1bs}). 
	For a particular instance, consider the following specific states and local filters(Eqs.(\ref{filters1},\ref{filters2})) involved in $\mathcal{N}_{n-star}^{(seq)}:$
	\begin{itemize}
		\item $\rho_1^{(\small{BD})}$ with $w_{11}$$=$$ 0.01$; $w_{12}$$=$$0.2456$; $w_{13}$$=$$ 0.639$; One can easily check that this state is Bell-CHSH local\cite{sh4}.
		\item $\rho_2^{\small{H}}$ with $p_{2}$$=$$0.042$ and $\theta_2$$=$$0.4585;$
		\item $\rho_3^{\small{H}}$ with $p_{3}$$=$$0.2169$ and $\theta_3$$=$$0.4585;$
		\item $\epsilon_2$$=$$0.99,$ $\epsilon_3$$=$$0.902,$ $\epsilon_4$$=$$0.998$ for edge parties $\textbf{A}_3,\textbf{A}_4$ respectively.
		\item $\epsilon_1^{(1)}$$=$$0.98,$ $\epsilon_1^{(2)}$$=$$0.5569,$  $\epsilon_1^{(3)}$$=$$0.98$ for central party $\textbf{A}_1.$
	\end{itemize}
	If above states are used in usual trilocal star network then $B_{3-star}$$=$$1.9082.$ So, trilocal inequality(Eq.(\ref{ineqs})) is not violated. But the same states when used in $\mathcal{N}_{3-star}^{(seq)}$ along with the specific filters mentioned above, then Eq.(\ref{up211f}) is satisfied as $B_{3-star}^{(seq)}$$=$$2.1039$ with approximately $37\%$ success probability. 
	\subsubsection{$\mathcal{N}_{n-star}^{(seq)}$ Involving Only One Hidden Nonlocal State}\label{hidex2}
\begin{center}
		\begin{figure}
			\includegraphics[width=3.4in]{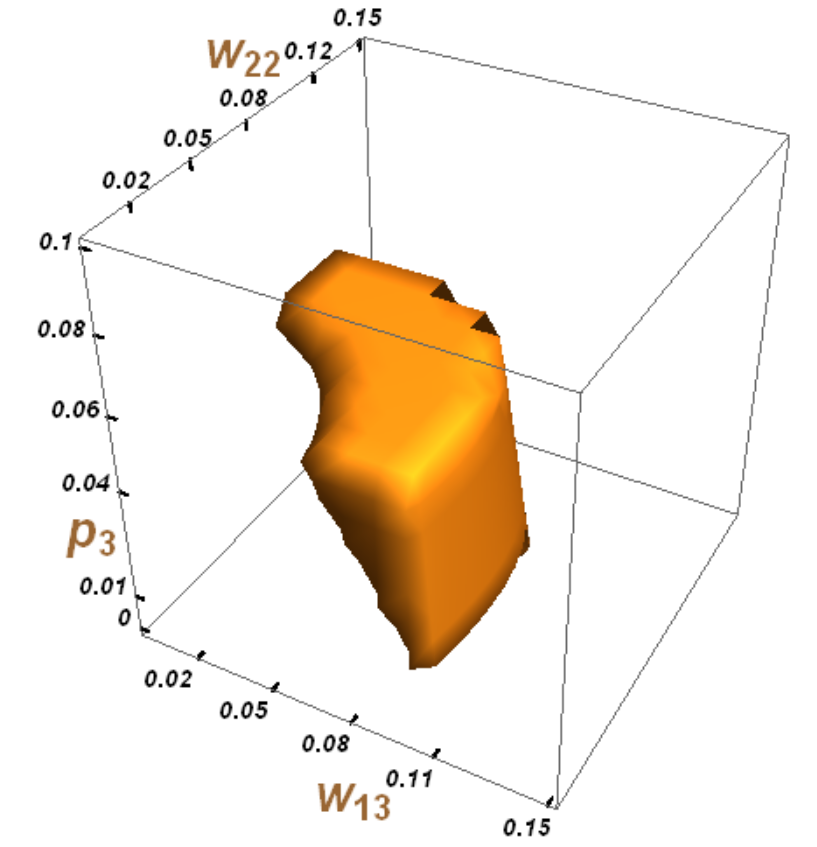} 
			\caption{\emph{Shaded region gives range of $w_{13},w_{22},p_3$ of $\rho_1^{(\small{BD})},\rho_2^{(\small{BD})}$ and $\rho_3^{\small{H}}$ respectively for which hidden non-trilocality is detected in $\mathcal{N}_{3-star}^{(seq)}$ with approximately $12\%$ probability of success. The specifications of the state parameters and filtering parameters(Eqs.(\ref{filters1},\ref{filters2})) are as follows: 
					$w_{11}$$=$$0.59,$ $w_{21}$$=$$0.57,$ $w_{12}$$=$$w_{23}$$=$$0,$
					$\epsilon_1^{(1)}$$=$$0.99,$ $\epsilon_1^{(2)}$$=$$0.99,$ $\epsilon_1^{(3)}$$=$$0.36,$ $\epsilon_4$$=$$0.36.$ }}
			\label{2b1g}
		\end{figure}
	\end{center}

	Let $\mathcal{S}_1,\mathcal{S}_2$ both distribute local Bell-Diagonal state, i.e., $\rho_1^{(\small{BD})},\rho_2^{(\small{BD})}$ both satisfy Eq.(\ref{belld2}). Let $\mathcal{S}_3$ distribute hidden nonlocal state $\rho_3^{\small{H}}$(Eq.(\ref{horodeck1})). For some local $\rho_1^{\small{BD}},\rho_2^{\small{BD}}$ and hidden nonlocal $\rho_3^{\small{H}}$ hidden non-trilocality is observed in the network under application of suitable local filters(see Fig.\ref{2b1g}). \\
	For a particular numerical example, consider the following specific states and local filters(Eqs.(\ref{filters1},\ref{filters2})):
	\begin{itemize}
		\item  $\rho_1^{(\small{BD})}$ with $w_{11}$$=$$ 0.485$; $w_{12}$$=$$0$; $w_{13}$$=$$ 0.512$; 
		\item $\rho_1^{(\small{BD})}$ with $w_{21}$$=$$ 0.532$; $w_{22}$$=$$0.465$; $w_{23}$$=$$ 0.$ $\rho_1^{(\small{BD})},\rho_2^{(\small{BD})}$ are both Bell-CHSH local.
		\item $\rho_3^{\small{H}}$ with $p_{3}$$=$$0.015$ and $\theta_3$$=$$0.785;$
		\item  $\epsilon_4$$=$$0.36$ for $\textbf{A}_4$.
		\item $\epsilon_1^{(1)}$$=$$0.99,$ $\epsilon_1^{(2)}$$=$$0.99,$ $\epsilon_1^{(3)}$$=$$0.37,$ for central party $\textbf{A}_1.$
	\end{itemize}
	$\textmd{For above specifications, }$$B_{3-star}$$=$$1.9983$ and $B_{3-star}^{(seq)}$$=$$2.00673$ with approximately $13\%$ probability of success. Hidden non-trilocality is thus observed. \\
	\par Based on above result together with Theorem.\ref{theo2}, following necessary criterion for detecting hidden non $n$-locality in $\mathcal{N}_{n-star}^{(seq)}$ becomes obvious: \\
	\textit{\textbf{Necessary Condition:} At least one of the sources must distribute a two-qubit state which is either  Bell-CHSH nonlocal or display hidden Bell-CHSH nonlocality under suitable filtering operations.} \\
	However above condition is not sufficient for the same. For example, let $\mathcal{S}_1$ distribute a hidden nonlocal state $\rho_1^{\small{H}}$ whereas each of the remaining sources distribute the following local state:
	\begin{equation}
		\varrho=\frac{1}{2}(|00\rangle\langle 00|+|00\rangle\langle 01|+|01\rangle\langle 00|+|01\rangle\langle 01|)
	\end{equation}
	Above state, being a two-qubit pure product state, gets transformed into another product state($\varrho^{'}$,say) after being subjected to local filters. Correlation tensor of $\varrho^{'}$ has only one non zero singular value. Consequently, by Theorem.\ref{theo1}, $B_{n-star}^{(seq)}$ turns out to be less than $2,$ indicating no violation of Eq.(\ref{ineqs}) and hence no detection of hidden non $n$-locality. \\
	\par So far we have considered $\mathcal{N}_{n-star}^{(seq)}$ where it is assumed that in the preparation phase of the protocol all the parties perform local filtering operations on its respective share of quantum states. But in practical situations such an assumption may not always be feasible. It may happen that all the parties do not perform local filtering operations. We explore next the possibility of generating non $n$-locality under such limitations.
	\section{Locally Accessible Hidden Non $n$-locality}\label{ress3}
	Before exploiting hidden non $n$-locality under the assumption that all the parties in $\mathcal{N}_{n-star}^{(seq)}$ do not perform local filters, let us first give a formal definition of \textit{locally accessible hidden non $n$-locality:}\\
	\textit{\textbf{Definition.2:}} \textit{In sequential n-local star network, with only a proper subset of the parties performing local filters in the preparation phase followed by each of the edge parties performing projective measurements in anyone of two possible directions and the central party performing only $n$-partite GHZ basis measurement in the measurement phase, if $n+1$-partite measurement correlations, restricted by the n-local constraint(Eq.(\ref{tr1})) in the network, cannot be decomposed in the form given by Eq.(\ref{tr2}), then such correlations are said to be locally accessible hidden non n-local correlations and the corresponding notion of nonlocality is referred to as locally accessible hidden non n-locality.}\\
	Clearly, exploiting locally accessible hidden non $n$-locality lowers down the restriction that all the parties must perform local filtering in $\mathcal{N}_{n-star}^{(seq)}.$ Precisely, any one of the following modifications of the preparation phase of $\mathcal{N}_{n-star}^{(seq)}$ can be made in this context:
	\begin{itemize}
		\item Only central party $\textbf{P}_1$ perform local filter $\mathfrak{F}_1$(Eq.(\ref{cenfil})) whereas none of the edge parties perform any operations in the first phase of the protocol.\\
		\item $\forall i$$=$$2,3,...,n+1,$ edge party $\textbf{P}_i$ performs local filter $\textbf{F}_i$ but the central party does not perform any operation in the first phase of the protocol.\\
		\item $\textbf{P}_1$ and $m$(say) number of edge parties with $1$$\leq$$m$$\leq$$n-1$ perform local filters.
	\end{itemize}
	Let us next illustrate locally accessible hidden non-trilocality for any of the above possible modifications in $\mathcal{N}_{3-star}^{(seq)}.$
	\subsection{Exploiting Locally Accessible Hidden Non trilocality}
\begin{center}
		\begin{figure}
			\begin{tabular}{cc}
				\subfloat[ ]{\includegraphics[trim = 0mm 0mm 0mm 0mm,clip,scale=0.37]{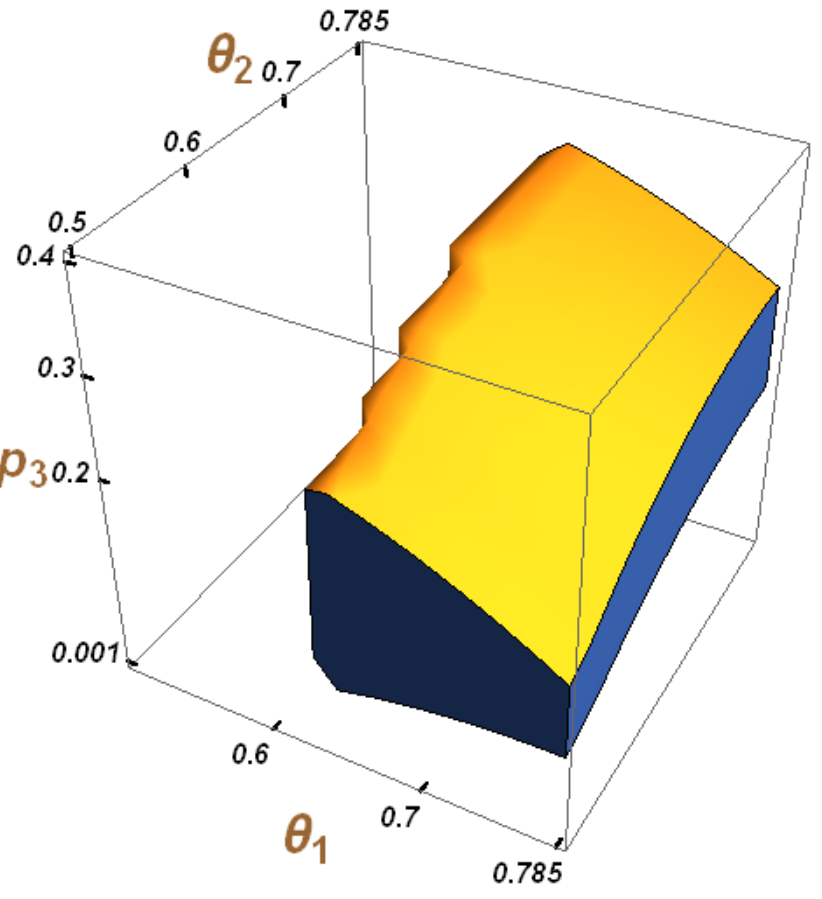}}\\
				\subfloat[]{\includegraphics[trim = 0mm 0mm 0mm 0mm,clip,scale=0.37]{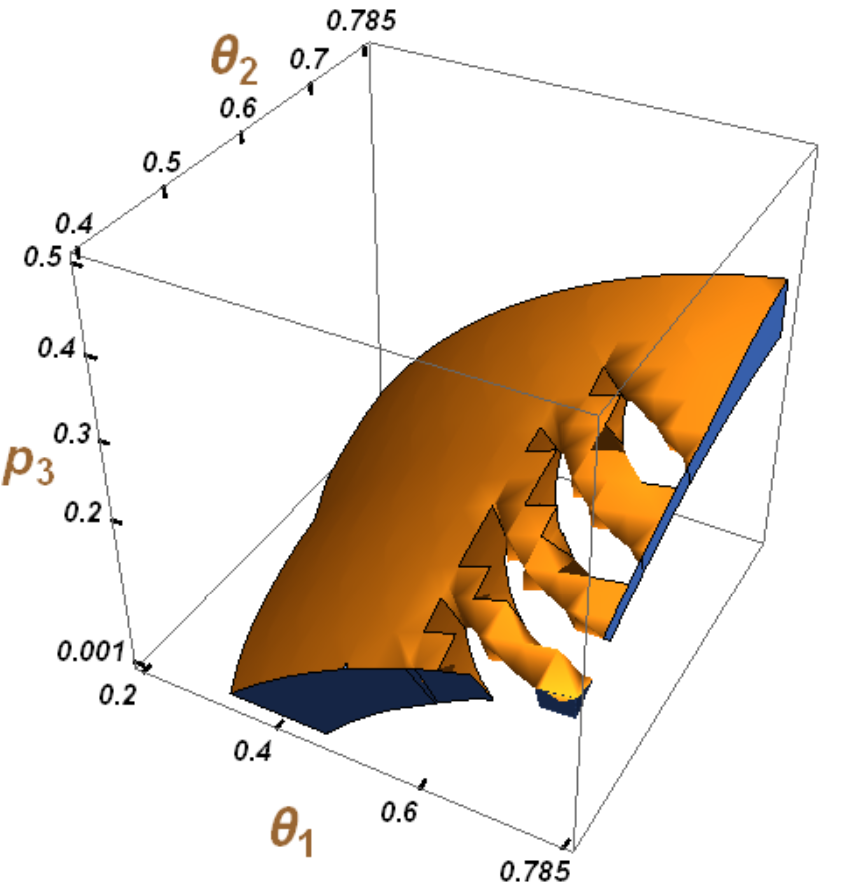}}&\\
				\subfloat[]{\includegraphics[trim = 0mm 0mm 0mm 0mm,clip,scale=0.37]{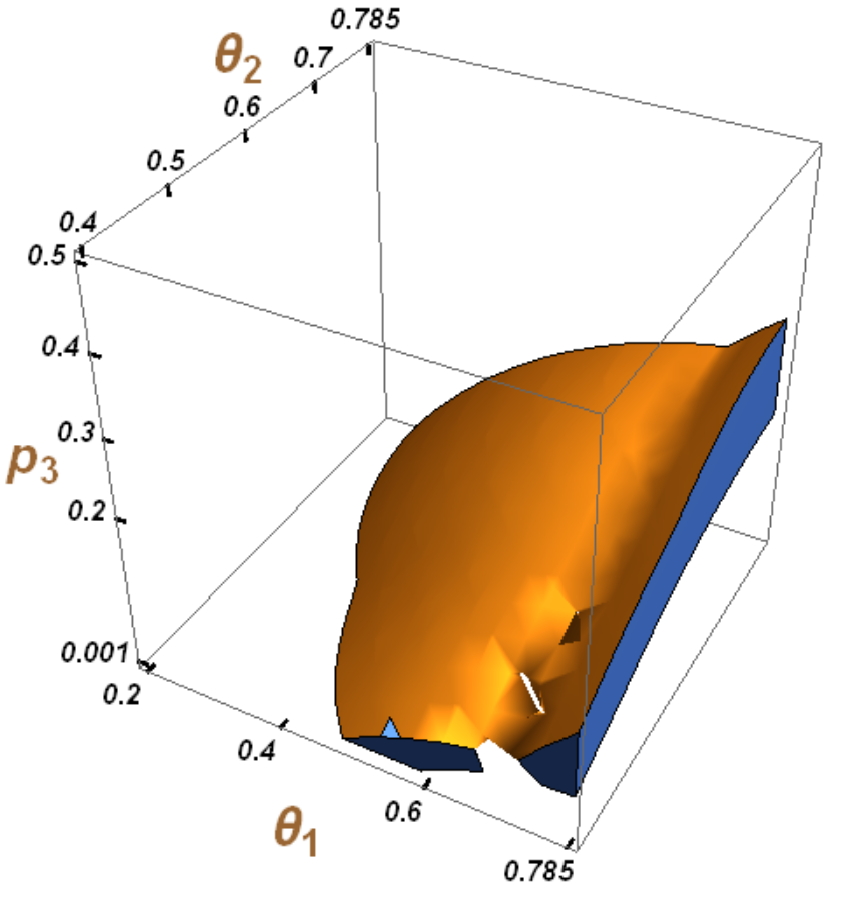}}\\
			\end{tabular}
			\caption{\emph{ Shaded region in each sub-figure gives the state parameters of $\rho_1^{(H)},\rho_2^{(H)},\rho_3^{(H)}$ for which hidden non-trilocality is locally accessible in $\mathcal{N}_{3-star}^{(seq)}$ for suitable filter parameters. One may refer to the $2,^{nd},3^{rd}$ columns of Table.\ref{table:taloc} for the filters specifications used here. State parameters fixed for plotting are as follows:$p_1$$=$$0.5,p_2$$=$$0.111$ for sub-fig.(i);$p_1$$=$$0.0833,p_2$$=$$0.10$ for sub-fig.(ii); $p_1$$=$$0.417,p_2$$=$$0.1$ for sub-fig.(iii) and  $\theta_3$$=$$\frac{\pi}{4}$ for all three sub-figures. }}
			\label{locac}
		\end{figure}
	\end{center}

	Let each of the sources $\mathcal{S}_i$ distribute $\rho_i^{(H)}.$ Let us first consider the case when only the central party applies local filters whereas none of $\textbf{P}_2,\textbf{P}_2,\textbf{P}_3$ perform any operation in the preparation phase of the protocol. Hidden non-trilocality is observed for suitable filtering operations by $\textbf{P}_1$(sub-fig.(i) in Fig.\ref{locac}).\\
	Hidden nontrilocal correlations are also generated if not the central party but all edge parties(sub-fig.(ii) in Fig.\ref{locac}) or at least one of the edge parties perform filtering operations(sub-fig.(iii) in Fig.\ref{locac}).\\
	Numerical instances for all these cases of obtaining locally accessible hidden non-trilocality are provided in Table.\ref{table:taloc}. 
	\begin{center}
		\begin{table}
			\caption{Instances of locally accessible hidden nontrilocality in $\mathcal{N}_{3-star}^{(seq)}$ are enlisted here. $2^{nd}$ column mentions the party(ies) applying filters in $1^{st}$ phase of the protocol.
				$4^{th}$ and $5^{th}$ columns together indicate that hidden non-trilocality is observed in $\mathcal{N}_{3-star}^{(seq)}.$  For all the instances states used in the network are $\rho_1^{(H)},\rho_2^{(H)},\rho_3^{(H)}.$ }
			\begin{center}
				\begin{tabular}{|c|c|c|c|c|c|}
					\hline
					\small{State}&Party&\small{Filter}&$B_{3-star}$&$B_{3-star}^{(seq)}$&\small{Success}\\
					\small{Parame-}&&\small{Parame-}&&&\small{Proba-}\\
					\small{ters}&&\small{ters}&&&\small{bility}\\
					\hline
					$(\theta_1,p_1)$$=$&$\textbf{P}_1$&$\epsilon_1^{(1)}$$=$$0.415$,&$1.8616$&$2.0697$&$31\%$\\
					$(0.65,0.5),$&&&&&(\small{approx})\\
					$(\theta_2,p_2)$$=$&&$\epsilon_1^{(2)}$$=$$0.98$&&&\\
					$(0.65,0.143),$&&&&&\\
					$(\theta_3,p_3)$$=$&&,$\epsilon_1^{(3)}$$=$$
					0.985$&&&\\
					$(\frac{\pi}{4},0.24)$&&&&&\\
					\hline
					$(\theta_1,p_1)$$=$&$\textbf{P}_2,\textbf{P}_3$&$\epsilon_2$$=$$0.457$,&$1.9489$&$2.0188$&$52\%$\\
					$(0.5,0.417),$&&&&&(\small{approx})\\
					$(\theta_2,p_2)$$=$&$\textbf{P}_4$&$\epsilon_3$$=$$0.88$&&&\\
					$(0.46,0.1),$&&&&&\\
					$(\theta_3,p_3)$$=$&&$\epsilon_4$$=$$0.98$
					&&&\\
					$(\frac{\pi}{4},0.12)$&&&&&\\
					\hline
					$(\theta_1,p_1)$$=$&$\textbf{P}_2$&$\epsilon_2$$=$$0.44$&$1.92249$&$2.0209$&$50\%$\\
					$(0.5,0.5),$&&&&&(\small{approx})\\
					$(\theta_2,p_2)$$=$&&&&&\\
					$(0.58,0.091),$&&&&&\\
					$(\theta_3,p_3)$$=$&&&&&\\
					$(\frac{\pi}{4},0.10)$&&&&&\\
					\hline
				\end{tabular}
			\end{center}
			\label{table:taloc}
		\end{table}
	\end{center}

	\section{Few Practical Perspectives}\label{ress4}
	Findings so far ensure existence of two-qubit states which when used in $\mathcal{N}_{n-star}^{(seq)},$ under application of suitable single-qubit filters, can generate non-trilocal correlations. Next we intend to discuss some utility of our findings from practical perspectives.
	\subsection{Robustness To Noise Enhancement}\label{noisy}
	In any quantum network, in idealistic situations, pure entangled states are supposed to be distributed among the distant parties(commonly known as \textit{nodes} \cite{reps}). But in practical situations, entanglement distribution undergoes unavoidable interaction with the environment. Entanglement thus gets transferred across noisy channels \cite{nie}. This leads to destruction of non-classical resources in the sense that the entangled states generated from the sources lose  potential to exhibit non $n$-locality. It thus becomes imperative to study any strategy for enhancing possibility of detecting non $n$-locality in noisy networks. We explore whether our sequential protocol can aid in reviving the potential of the noisy entangled states to exhibit non $n$-locality. Applying suitable local filters turns out to be effective in increasing resistance to noise of the $n$-local inequality for detecting non $n$-local correlations in $\mathcal{N}_{n-star}^{(seq)}$. This in turn points out the utility of the protocol via which we can  retrieve power of some noisy entangled states to generate non $n$-local correlations. \\
	\par We have observed that for some noisy communication through channels, non-trilocality can be detected over a wider range of noise parameters in $\mathcal{N}_{3-star}^{(seq)}$ compared to that in $\mathcal{N}_{3-star}.$ Such observations will justify our claim that $\mathcal{N}_{3-star}^{(seq)}$ can be used in the above mentioned context.
	\subsubsection{Communication Through Amplitude Damping Channel\cite{nie}}
\begin{center}
	\begin{figure}
		\includegraphics[width=3.2in]{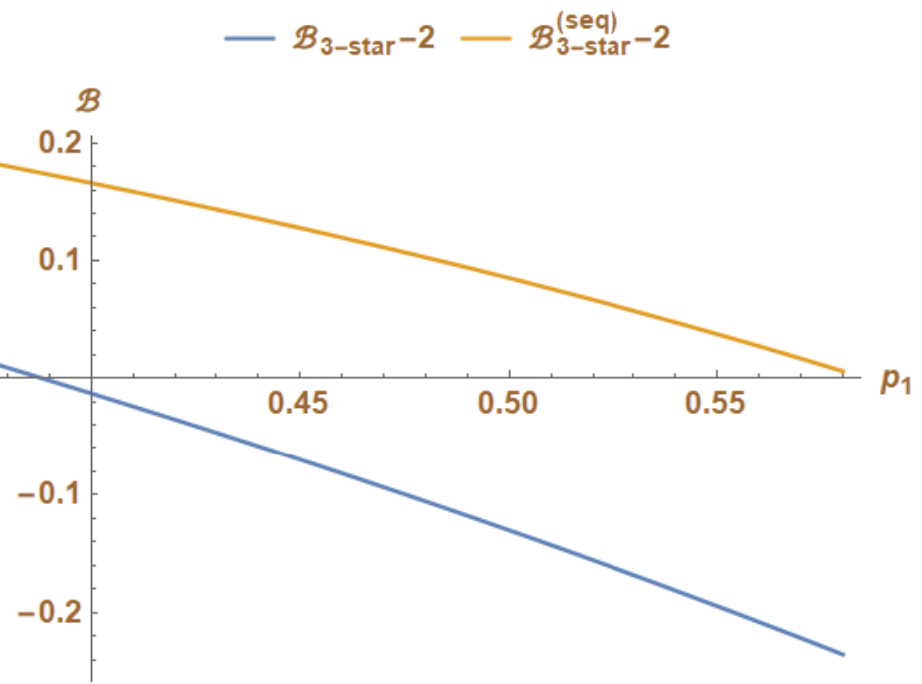} \\
		\caption{\emph{Plotting curves $\mathcal{B}_{3-star}^{(seq)}$$-$$2$ and $\mathcal{B}_{3-star}$$-$$2$. Specifications of the state, noise and filtering parameters are as mentioned in the main text. Comparison of the two curves point out existence of range of noise parameter $p_1$ for which hidden non trilocality is detected. }}
		\label{fignewa2}
	\end{figure}
\end{center}

	Let each of $\mathcal{S}_1,\mathcal{S}_2$ and $\mathcal{S}_3$  distribute an identical copy of pure entangled state:
	\begin{eqnarray}\label{pure1}
		|\Psi(\theta\rangle)&=&\sin\theta|01\rangle+\cos\theta |10\rangle,\,\theta\in(0,\frac{\pi}{4})
	\end{eqnarray}
	Let each of the qubits of $|\Psi(\theta\rangle)$ get distributed through identical amplitude damping channels\cite{nie} characterized by damping parameter $p_i(i$$=$$1,2,3).$ The mixed entangled states thus distributed in the network are given by $\rho_i^{(H)}$(Eq.(\ref{horodeck1})).
	It is observed that there exists range of the noise parameters $(p_1,p_2,p_3)$ for which non-trilocality can be exploited in $\mathcal{N}_{3-star}^{(seq)},$ under effect of suitable filtering operations in contrast to $\mathcal{N}_{3-star}$ where trilocal inequality(Eq.(\ref{ineqs})) is not violated. \\
	For example, consider the followings: $|\Psi(\frac{\pi}{5})\rangle$ and amplitude damping channels with $(p_2,p_3)$$=$$(0.12,0.2)$ and $p_1$$\in$$(0.377,0.575).$ With these specifications, $\mathcal{B}_{3-star}$$\leq$$2.$ Now, for  $(\epsilon_1^{(1)},\epsilon_1^{(2)},\epsilon_1^{(3)})$$=$$(0.49,0.99,0.99)$ and $\epsilon_2$$=$$1,$$\epsilon_3$$=$$\epsilon_4$$=$$0.98,$ $\mathcal{B}_{3-star}^{(seq)}$$>$$2.$  $(0.377,0.575)$ is thus the enhanced range of noise parameter for detecting hidden non-trilocal correlations in star network(see Fig.\ref{fignewa2}).\\
	
	\par However, such recovery of the potential of states to exhibit non-trilocality, after being communicated through noisy channels, is not always possible for all noisy channels via our protocol. A strong no-go result is obtained in the sense that sometimes recovery of non $n$-locality(upto detection via $n$-local inequality violation) becomes impossible when at least one source distributes two-qubit product state whereas rest of the qubits get distributed from the sources along any noisy channel:
	\begin{theorem}\label{theo22}
		\textit{In $\mathcal{N}_{n-star}^{(seq)}$ involving at least one two-qubit product state, $n$-local inequality cannot be violated. }
	\end{theorem}
	\textit{Proof:} See Appendix.C.\\
	\subsection{Detecting Source Entanglement}
	Let us consider a sequential $n$-local star network where we only have the information that each of $n$ independent sources distribute a two-qubit state. Observation of violation of $n$-local inequality will help us to detect entanglement of the sources. In this context, following no-go result holds:
	\begin{theorem}\label{theo3}
		\textit{In $\mathcal{N}_{n-star}^{(seq)},$ violation of the $n$-local inequality is impossible if each of the sources distribute a separable two-qubit state.}
	\end{theorem}
	\textit{Proof:} The proof follows directly from the findings of \cite{bilocent}, strategy used to prove Theorem.\ref{theo1} along with the fact that any separable two-qubit state remains separable after being subjected to local filtering operations.\\
	\par So, if violation of Eq.(\ref{ineqs}) is observed in $\mathcal{N}_{n-star}^{(seq)},$ above theorem ensures that at least one of the sources is entangled. Hence, presence of entanglement is detected across the network. However, it cannot give any information regarding the number of entangled sources in $\mathcal{N}_{n-star}^{(seq)}.$
	\section{Sequential Network With Non-Separable Multi-Qubit Filters}\label{ress6}
	In $\mathcal{N}_{n-star}^{(seq)}$ considered so far, each of the qubits is subjected to single-qubit filtering operations(Eq.(\ref{cenfil})). In such a scenario the central party($\textbf{P}_1$) thus performs separable filtering operations(Eq.(\ref{cenfil})) over each of the qubits that it receives from the sources. However, unlike the edge parties, $\textbf{P}_1$ has access to joint state of multiple qubits. One can thus consider a protocol where $\textbf{P}_1$ perform non-separable filtering operations on the joint state of the multi qubits that it has access to. In this section, we consider such a sequential set up involving $n$-local star network. We discuss the scenario below.
\subsection{Setting Up The Sequential Protocol}
This scenario is almost similar to the previous one with only modifications coming in the preparatory phase on the part of the central party. So explicitly we have:
\begin{itemize}
\item \textit{Preparatory Phase:} It is similar to the preparatory phase in $\mathcal{N}_{n-star}^{(seq)}$ with only alteration in the part of the central party. $\textbf{P}_1$  performs a non-separable $m$-qubits($2$$\leq$$m$$\leq$$n$) filtering operation($\mathcal{F}_{NS}$,say) on the joint state of $n$ qubits that it receives from $n$ independent sources(one from each source). Being filtering operation, $\mathcal{F}_{NS}$ satisfies $\mathcal{F}_{NS}^{\dagger}\cdot \mathcal{F}_{NS}$$\leq$$\mathbb{I}_{2^{\otimes n}}$ and cannot be decomposed into single-qubit filters, i.e., inexplicable in the form prescribed by Eq.(\ref{cenfil}). \\
After all the parties perform filtering, the filtered state 
		shared across network takes the form:
		\begin{eqnarray}\label{fil7n}
			\rho^{(f)}&=&\textbf{N}(\mathcal{F}_{NS}\otimes_{i=2}^{n+1}  \textbf{F}_{i})\rho_{\small{initial}}(\mathcal{F}_{NS}\otimes_{i=2}^{n+1}  \textbf{F}_{i})^{\dagger}\nonumber\\
			\textmd{where}\,\textbf{N}&=&\frac{1}{\textmd{Tr}((\mathcal{F}_{NS}\otimes_{i=2}^{n+1}  \textbf{F}_{i})\rho_{\small{initial}}(\mathcal{F}_{NS}\otimes_{i=2}^{n+1}  \textbf{F}_{i})^{\dagger})}
		\end{eqnarray}
		\item \textit{Measurement Phase:} Exactly same as that in $\mathcal{N}_{n-star}^{(seq)}.$
	\end{itemize}
	Let the sequential star network scenario with central party performing non-separable filters be denoted as $\mathbf{N}_{n-star}^{(seq)}.$ Let us next illustrate few instances of hidden non $n$-locality obtained in $\mathbf{N}_{n-star}^{(seq)}.$ We consider $n$$=$$3$ for our purpose.
	\subsection{Illustrating Hidden Non Trilocality}
\begin{center}
		\begin{figure}[H]
			\includegraphics[width=3.4in]{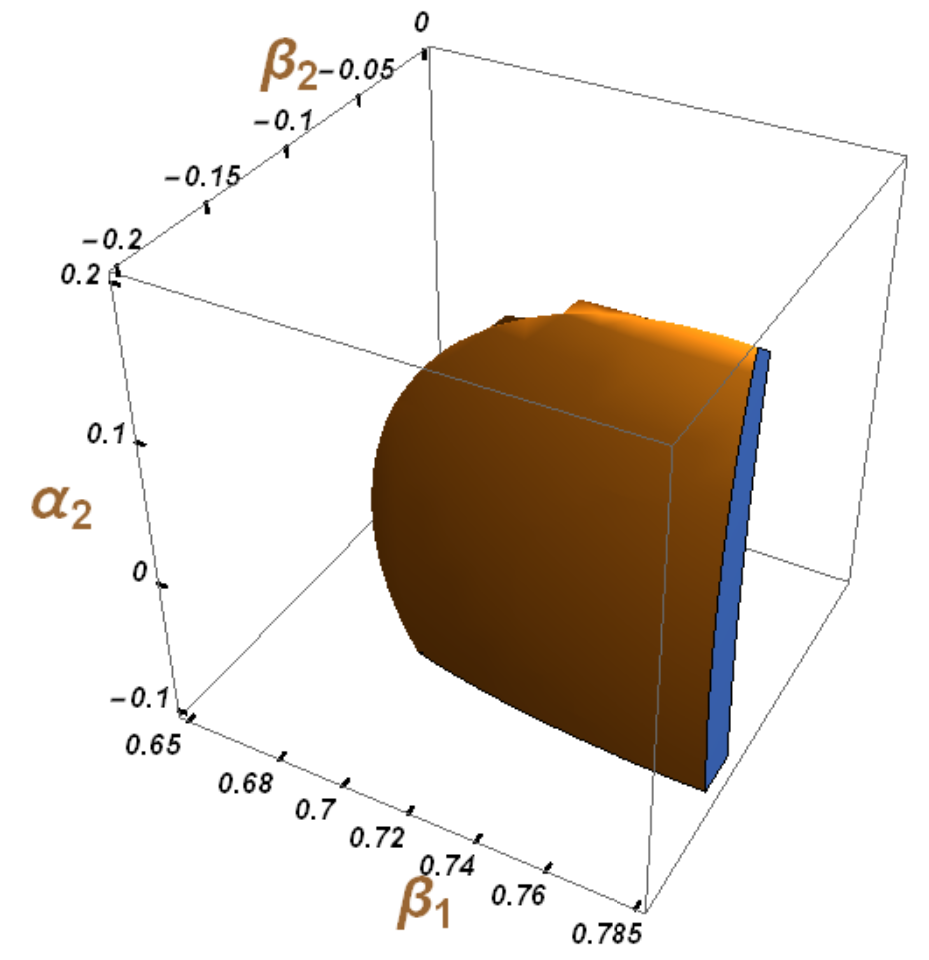} 
			\caption{\emph{Shaded region indicates parameters of pure entangled states(Eq.(\ref{pure1})) and parameter corresponding to $3$-qubit filtering operation $\mathcal{F}_{NS}$(together with $\alpha_1$$=$$0.27143$ in Eq.(\ref{pure3}) for which hidden non-trilocality is detected via $\mathbf{N}_{3-star}^{(seq)}$ with approximately $9\%$ probability of success.}}
			\label{multfilte2}
		\end{figure}
	\end{center}

As already discussed before, it is impossible to observe violation of $n$-local inequality(Eq.(\ref{ineqs})) in usual $n$-local star network with at least one source distributing a two-qubit product state. Let us consider $\mathbf{N}_{3-star}^{(seq)}$ with the following specifications:
	\begin{itemize}
		\item  $\mathcal{S}_i$ distribute two-qubit pure entangled state $	|\psi_i\rangle\langle \psi_i|$ where:
		\begin{eqnarray}\label{pure1}
			|\psi_i\rangle&=&\sin \beta_i |01\rangle+\cos\beta_i|10\rangle,	\\
			&&\beta_i\in[-\frac{\pi}{4},\frac{\pi}{4}]\,\forall i=1,2\nonumber
		\end{eqnarray}
		\item  $\mathcal{S}_3$ distribute two-qubit pure product state:
		\begin{eqnarray}\label{pure2}
			\rho_3^{(prod)}&=&\frac{1}{4}(\sum_{i,j,k,l=0}^1|ij\rangle\langle kl|)
		\end{eqnarray}
		\item	 $\mathbf{P}_1$ perform the following $3$-qubit non-separable filtering operation:
		\begin{eqnarray}\label{pure3}
			\mathcal{F}_{NS}&=&\alpha_1 |000\rangle\langle 000|+|001\rangle\langle 001|+\alpha_1 |010\rangle\langle 010|+\nonumber\\
			&&|011\rangle\langle 011|+\alpha_1|100\rangle\langle 100|+|101\rangle\langle 101|+\nonumber\\
			&&\alpha_1\cos\frac{\alpha_2}{2}|110\rangle\langle 110|+\sin\frac{\alpha_2}{2}|111\rangle\langle 110|\nonumber\\
			&&-\alpha_1\sin\frac{\alpha_2}{2}|110\rangle\langle 111|+\cos\frac{\alpha_2}{2}|111\rangle\langle 111|
		\end{eqnarray}
		\item None of the edge parties perform any filtering operation.
	\end{itemize}
For suitable filtering parameters(Eq.(\ref{pure3})) and measurement settings, non-trilocal correlations are generated in $\mathbf{N}_{3-star}^{(seq)}$(see Fig.\ref{multfilte2}).\\
	\par
\begin{center}
		\begin{figure}[H]
			\includegraphics[width=3.4in]{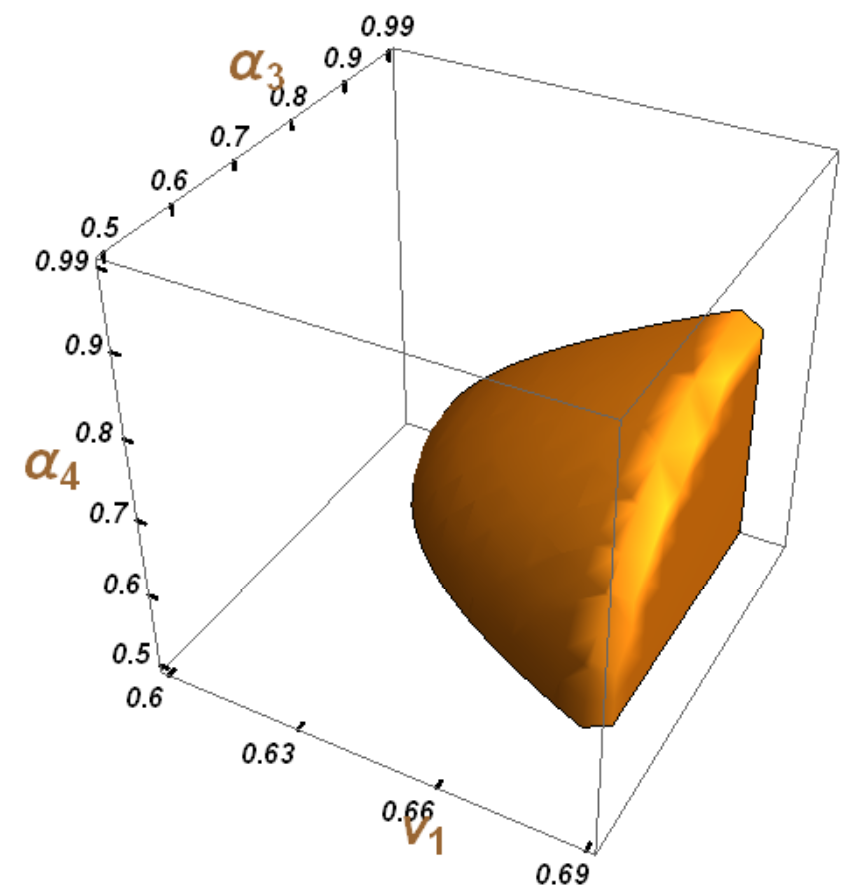} 
			\caption{\emph{Shaded region indicates range of noise parameter $v_1$ and the  parameters characterizing filter $\mathcal{F}_{NS}$(Eq.(\ref{2fil1})) $(\alpha_3,\alpha_4)$ for which non-trilocal correlations are obtained in $\mathbf{N}_{3-star}^{(seq)}$  with approximately $25\%$ probability of success. Other fixed state parameters involved here are $p_3$$=$$0.1,$ $\theta_3$$=$$\frac{\pi}{7}$ and $v_2$$=$$0.69.$ For the same states, non-trilocality cannot be detected in usual trilocal star network. So shaded region indicates exploitation of hidden non-trilocality.}}
			\label{multfilt1}
		\end{figure}
	\end{center}

 Let us next consider an example where the sources distribute mixed entangled states.
Consider Werner class of two-qubit states\cite{Wer}:
\begin{eqnarray}\label{wer1}
	\rho_i^{(Wer)}&=&\frac{(1-v_i)}{4}\mathbb{I}_{4\times 4}+v_i(|01\rangle\langle 01|+|10\rangle\langle 10|\nonumber\\
	&&-(|01\rangle\langle 10|+|10\rangle\langle 01|))\ , v_i\in[0,1]
\end{eqnarray}
Let each of $\mathcal{S}_1$ and $\mathcal{S}_2$ distribute Bell-CHSH local Werner state: $\rho_1^{(Wer)}$ and $\rho_2^{(Wer)}$(Eq.(\ref{wer1}) with $v_1,v_2$$\leq$$\frac{1}{\sqrt{2}})$. Let $\mathcal{S}_3$ distribute $\rho_3^{(H)}$(Eq.(\ref{horodeck1})). When these states are used in usual trilocal star network, violation of $n$-local inequality(Eq.(\ref{ineqs})) is not observed.\\
	Let these three states be now used in $\mathbf{N}_{n-star}^{(seq)}.$ Let $\mathbf{P}_1$ perform following $2$-qubit filtering operation:
	\begin{eqnarray}\label{2fil1}
		\mathcal{F}_{NS}&=&0.9779\alpha_3\alpha_4 |00\rangle\langle 00|-\nonumber\\
		&&0.2089\|11\rangle\langle 00|+0.9779\alpha_3|01\rangle\langle 01|+\nonumber\\
		&& 0.2089\alpha_4|10\rangle\langle 01|-0.2089\alpha_3 |01\rangle\langle 10|+\nonumber\\
		&&0.9779\alpha_4 |10\rangle\langle 10|+0.2089\alpha_3\alpha_4 |00\rangle\langle 11|\nonumber\\
		&&+0.9779|11\rangle\langle 11|
	\end{eqnarray}
	Above filter operation(Eq.(\ref{2fil1})) is applied on joint state of two of the three qubits(say $2^{nd},3^{rd}$ qubits) that $\textbf{P}_1$ has received. One can see that such a filter is non-separable. Let none of the edge parties perform any filtering operation. 
	For some state parameters $(v_1,v_2,p_3),$ non-trilocal correlations are generated in $\mathbf{N}_{n-star}^{(seq)}$(see Fig.\ref{multfilt1}). For numerical instances see Table.\ref{table:talocsep}.
	\subsection{Comparison Between $\mathcal{N}_{n-star}^{(seq)}$ And $\mathbf{N}_{n-star}^{(seq)}$}
	Theorem.\ref{theo22} ensures impossibility of detecting hidden non of $n$-locality in $\mathcal{N}_{n-star}^{(seq)}$ when at least one of the independent sources distributes a two-qubit product state. However, such a no-go result does not hold good in $\mathbf{N}_{n-star}^{(seq)}.$ This in turn points out the effectiveness of non-separable multi-qubit filters in sequential set-up. We next justify our comment  with some examples.\\
	Consider $\mathbf{N}_{3-star}^{(seq)}$ involving two non identical copies of Werner states $\rho_1^{(Wer)}$ and $\rho_3^{(Wer)}$(Eq.(\ref{wer1}). Let $\mathcal{S}_2$ distribute $\rho_2^{(prod)}$(Eq.(\ref{pure2})).
	
\begin{center}
		\begin{figure}
			\includegraphics[width=3.2in]{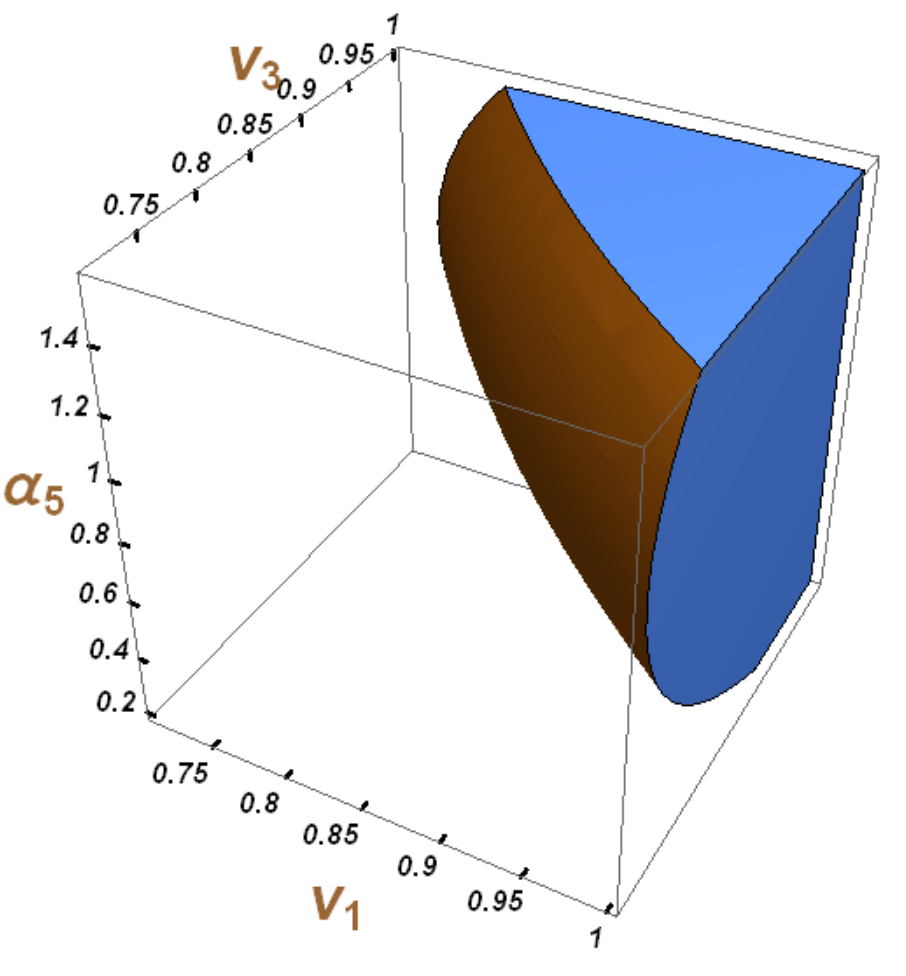} 
			\caption{\emph{Shaded region indicates range of noise parameters $v_1,v_3$ and the  filter parameter corresponding to filter $\mathcal{F}_{NS}$(Eq.(\ref{2fil4}) with $\alpha_6$$=$$0.$) $\alpha_5$ for which hidden non-trilocality is exploited with approximately $40\%$ probability of success(for suitable measurement settings in measurement phase) via our protocol. }}
			\label{multfilt2}
		\end{figure}
	\end{center}

	Let $\mathbf{P}_1$ perform the following filtering operation:
	\begin{eqnarray}\label{2fil4}
		\mathcal{F}_{NS}&=&(0.8|0\rangle\langle 0|+|1\rangle\langle 1|)\otimes G\,\, \textmd{where}\nonumber\\
		G&=&0.64\cos\frac{\alpha_5}{2}\cos\frac{\alpha_6}{2} |00\rangle\langle 00|+0.8\sin\frac{\alpha_5}{2}\nonumber\\
		&&\sin\frac{\alpha_6}{2} |01\rangle\langle 00|-0.8\cos\frac{\alpha_5}{2}\sin\frac{\alpha_6}{2} |10\rangle\langle 00|\nonumber\\
		&&-\sin\frac{\alpha_5}{2}\cos\frac{\alpha_6}{2} |11\rangle\langle 00|+0.64\sin\frac{\alpha_5}{2}\nonumber\\
		&&\sin\frac{\alpha_6}{2} |00\rangle\langle 01|+0.8\cos\frac{\alpha_5}{2}\cos\frac{\alpha_6}{2} |01\rangle\langle 01|+\nonumber\\
		&&0.8\sin\frac{\alpha_5}{2}\cos\frac{\alpha_6}{2} |10\rangle\langle 01|+\cos\frac{\alpha_5}{2}\sin\frac{\alpha_6}{2} |11\rangle\langle 01|\nonumber\\
		&&+0.64 \cos\frac{\alpha_5}{2}\sin\frac{\alpha_6}{2} |00\rangle\langle 10|-0.8\sin\frac{\alpha_5}{2}\nonumber\\
		&&\cos\frac{\alpha_6}{2} |01\rangle\langle 10|+0.8\cos\frac{\alpha_5}{2}\cos\frac{\alpha_6}{2} |10\rangle\langle 10|-\nonumber\\
		&&\sin\frac{\alpha_5}{2}\sin\frac{\alpha_6}{2} |11\rangle\langle 10|+0.64\sin\frac{\alpha_5}{2}\cos\frac{\alpha_6}{2} |00\rangle\langle 11|\nonumber\\
		&&-0.8 \cos\frac{\alpha_5}{2}\sin\frac{\alpha_6}{2} |01\rangle\langle 11|-0.8\sin\frac{\alpha_5}{2}\nonumber\\
		&&\sin\frac{\alpha_6}{2} |10\rangle\langle 11|+\cos\frac{\alpha_5}{2}\cos\frac{\alpha_6}{2} |11\rangle\langle 11|
	\end{eqnarray}
	Clearly above filtering operation(Eq.(\ref{2fil4})) is $2$-qubit non-separable.
	For some range of visibility parameters $v_1,v_3,$ hidden non-trilocality is detected in $\mathbf{N}_{n-star}^{(seq)}$(see Fig.\ref{multfilt2}). For particular numerical instance, one may refer to Table.\ref{table:talocsep}.
	This in turn points out utility of $\mathbf{N}_{n-star}^{(seq)}$ in enhancing robustness to noise of the noisy entangled states for generating hidden non-trilocal correlations. \\
	\begin{center}
		\begin{table}
			\caption{Instances of hidden non-trilocality exploited in $\mathbf{N}_{3-star}^{(seq)}$ are enlisted here. Only the central party($\textbf{P}_1$) applies filters in $1^{st}$ phase of the protocol.
				$3^{rd}$ and $4^{th}$ columns together indicate that hidden non-trilocality is observed in $\mathbf{N}_{3-star}^{(seq)}.$ Approx probability of success($N^{-1}$) is mentioned in the last column.}
			\begin{center}
				\begin{tabular}{|c|c|c|c|c|}
					\hline
					\small{State}&\small{Filter}&$B_{3-star}$&$B_{3-star}^{(seq)}$&$N^{-1}$\\
					\small{Parame-}&\small{Parame-}&&in $\mathbf{N}_{3-star}^{(seq)}.$ &\\
					\small{ters}&\small{ters}&&&\\
					\hline
					$(\beta_1,\beta_2)$$=$&$\alpha_1$$=$$0.26$&$1.31468$& $2.00716$&$9\%$\\
					$(0.785,-0.144)$	&$\alpha_2$$=$$0.173$&&&(\small{approx})\\
								\hline
					$(\theta_3,p_3)$$=$&$\alpha_3$$=$$0.88$,&$1.9980$&$2.0408$&$49\%$\\
					$(\frac{\pi}{7},0.1),$&\small{in Eq.(\ref{pure3})}&&&\\
					$v_1$$=$$0.68,$&$\alpha_4$$=$$0.67$&&&(\small{approx})\\
					$v_2$$=$$0.69$&&&&\\
					\hline
					$v_1$$=$$0.85,$&$\alpha_5$$=$$1.1$&$1.80164$&$2.0056$&$52\%$\\
					$v_3$$=$$0.86$&$\alpha_6$$=$$-\frac{\pi}{2}$&&&(\small{approx})\\
					\hline
				\end{tabular}
			\end{center}
			\label{table:talocsep}
		\end{table}
	\end{center}
\section{Concluding Remarks}\label{ress7}
With progress in development of scalable quantum networks, quantum network correlations are nowadays used as a resource in different information processing tasks. Present study has given a protocol by modifying existing star-shaped $n$-local networks so as to incorporate stage of local filtering operations by the parties before they perform local measurements on their respective qubits. Such a sequential set-up turned out to be advantageous in context of generating non $n$-local correlations in some cases. Instances of hidden non trilocality have been provided in this regard. However, several limitations of such  sequential star-shaped networks are also obtained which have been put in forms of no-go results. For example, hidden non $n$-locality cannot be detected if each of the source distributes a two-qubit state that cannot reveal any hidden Bell-CHSH nonlocality\cite{rajs}. Examples of non-separable $3$-qubit filters have been provided to prove sequential trilocal networks emerge more efficient to detect non trilocality when central party performs non-separable filters instead of separable form of $3$-qubit filters. To this end it may be noted that few discrete classes of non-separable $3$-qubit filters have been provided here. It will be better if one can provide a comprehensive study of sequential $n$-local networks involving $n$-qubit non-separable filtering operations. \\
\par To detect hidden Bell-CHSH  nonlocality of a two-qubit state, a necessary and sufficient criterion exists\cite{rajs}. It will be interesting to construct a similar detection criterion for hidden non $n$-locality. Any such criterion will help in characterizing entire two-qubit state space in this context. Partial characterization of sources in context of exhibiting non $n$-locality has been provided here. It will be interesting if one can completely characterize source entanglement from the perspective of detecting hidden non $n$-locality. 

	\section*{Appendix.A}\label{theoa20}
	\textit{Proof of Theorem.\ref{theo1}:}
	Consider a  two-qubit state $\rho_{1,i}$ generated from source $\mathcal{S}_i.$ In the preparation phase of the protocol, both the qubits of $\rho_{1,i}$ are subjected to local filtering operations:
	\begin{itemize}
		\item $\textbf{F}_i$ by edge party $\textbf{P}_i$
		\item $\textbf{F}_1^{(i)}$ by central party $\textbf{P}_1.$
	\end{itemize}
	At the end of the preparation phase the state $\rho^{(f)}$ (Eq.(\ref{fil7})) shared across the entire network is thus given by:
	\begin{eqnarray}\label{fil7i}
		\rho^{(f)}&=&\textbf{N}(\mathfrak{F}\otimes_{i=2}^{n+1}  \textbf{F}_{i})\rho_{\small{initial}}(\mathfrak{F}\otimes_{j=2}^{n+1}  \textbf{F}_{j})^{\dagger}\nonumber\\
		&=&\frac{1}{\Pi_{i=1}^n C_i}\otimes_{i=1}^n(\textbf{F}_i\otimes \textbf{F}_1^{(i)}.\rho_{1,i}.(\textbf{F}_i\otimes \textbf{F}_1^{(i)})^{\dagger} )\nonumber\\
		&=&\otimes_{i=1}^n\rho_{1,i}^{(f)}\,\,\textmd{where}, \\
		\rho_{1,i}^{(f)}&=&\frac{1}{\Pi_{i=1}^n C_i}\otimes_{i=1}^n\textbf{F}_i\otimes \textbf{F}_1^{(i)}.\rho_{1,i}.(\textbf{F}_i\otimes \textbf{F}_1^{(i)})^{\dagger},\nonumber\\
		C_i&=&\textmd{Tr}[\textbf{F}_i\otimes \textbf{F}_1^{(i)}.\rho_i.(\textbf{F}_i\otimes \textbf{F}_1^{(i)})^{\dagger}]
	\end{eqnarray}
	Local measurements are performed on $\rho^{(f)}$ in the measurement phase. The separable structure(Eq.(\ref{fil7i})) of $\rho^{(f)}$ implies that normalized post-selected two-qubit states $\rho_{1,1}^{(f)},\rho_{1,2}^{(f)},...,\rho_{1,n}^{(f)}$ are involved in the measurement phase of the protocol. The  procedure of maximizing the value of the $n$-local inequality(Eq.(\ref{ineqs})) with respect to the measurement parameters will be same as that used in \cite{bilo5}. Upper bound of Eq.(\ref{ineqs}) will thus be given by the largest two singular values $W_{i,1}^{(f)}$$\geq$$W_{i,2}^{(f)}$ of the correlation tensors of $\rho_{1,i}^{(f)}(i$$=$$1,2,...,n),$ i.e., $\cdot B_{n-star}^{(seq)}$ where:
	\begin{eqnarray}
		B_{n-star}^{(seq)}=2\sqrt{\Pi_{i=1}^n(W_{i,1}^{(f)})^{\frac{2}{n}}+\Pi_{i=1}^n(W_{i,2}^{(f)})^{\frac{2}{n}}}
	\end{eqnarray}
	The theorem is proved.\\
	\section*{Appendix.B}\label{theoa21}
	\textit{Proof of Theorem.\ref{theo2}:} By given condition of the theorem, each of $\rho_i$ is Bell-CHSH local upto application of local filters. So, $\forall i,$ $\rho_i^{(f)}$ is Bell-CHSH local\cite{horo}. Hence, 
	\begin{equation}\label{fil7iii}
		(W_{i1}^{(f)})^2+ (W_{i2}^{(f)})^2\leq 1,\, i=1,2,...,n
	\end{equation}
	By Eq.(\ref{fil7ii}), we get, 
	\begin{eqnarray}\label{fil7iv}
		B_{n-star}^{(seq)}&\leq&2\sqrt{\frac{\sum_{i=1}^n  (W_{i1}^{(f)})^2+\sum_{i=1}^n  (W_{i2}^{(f)})^2}{n}}\,(By\, A.M\geq G.M.)\nonumber\\
		&=&2\sqrt{\frac{\sum_{i=1}^n ((W_{i1}^{(f)})^2+ (W_{i2}^{(f)})^2)}{n}}\nonumber\\
		&\leq& 2\,(By,\, Eq.(\ref{fil7iii}))
	\end{eqnarray}
	\section*{Appendix.C}\label{theoa22}
	\textit{Proof of Theorem.\ref{theo22}:} Proof of the theorem is straightforward based on the fact that any two-qubit product state remains a product state only after its both qubits are subjected to single qubit local filters. \\
	W.L.O.G., let $\rho_1$ be a two-qubit product state. Its correlation tensor matrix is diagonal with only one non zero entry($t$ say with $|t|$$\leq$$1$). Let $\rho_1^{(f)}$ denote the state after being subjected to filtering. $\rho_1^{(f)}$ still remains a product state. So the correlation tensor matrix of $\rho_1^{(f)}$ is diagonal with only one non zero entry($t^{(f)}$ say with $|t^{(f)}|$$\leq$$1$). \\
	By Theorem.\ref{theo1}, when $\rho_1$ is used with other arbitrary two-qubit states $\rho_2,...,\rho_n$ in our sequential network, then we get $B_{n-star}^{(seq)}$$\leq$$2.$ Hence no violation of $n$-star inequality is obtained in $\mathcal{N}_{n-star}^{(seq)}.$ Proved.
		
\end{document}